\documentstyle[preprint,eqsecnum,cite,aps]{revtex}

\begin{document}
\draft \title{On the Nonrelativistic Limit of the $\varphi ^4$ Theory
in 2+1 Dimensions \cite{byline}} \author{M. Gomes,
J. M. C. Malbouisson\cite{Ma}, and A. J. da Silva} \address{Instituto
de F\'\i sica, Universidade de S\~ao Paulo, Caixa Postal 66318,\\
05315--970, S\~ao Paulo, SP, Brazil.}  

\maketitle

\begin{abstract}
We study the nonrelativistic limit of the quantum theory of a real
scalar field with quartic self-interaction. The two body scattering
amplitude is written in such way as to separate the contributions of
high and low energy intermediary states. From this result and the two
loop computation of the self energy correction, we determine an effective
nonrelativistic action.
\end{abstract}

\narrowtext
\section{INTRODUCTION}

It is generally believed that nonrelativistic field theories can be obtained
from corresponding relativistic ones as appropriated limits for low momenta.
Earlier attempts to the quest of a reliable scheme of nonrelativistic
approximation have been based on canonical transformations over the
Lagrangians$^{\citen{fol}}$. More recently, there has been proposals of
construction of effective Lagrangians by amending the nonrelativistic
theories with other interaction terms, representing the effect of the
integration of the relativistic degrees of freedom, inspired in the
renormalization group spirit$^{\citen{lep}}$. In any case, the goodness of any
conceivable approximation will certainly rely on how and how much of the
high energy part of the Hilbert space is considered to influence the
low energy sector of the theory.

Nonrelativistic field theories in 2 + 1 dimensions present many interesting aspects. The simplest of
them, $\lambda \phi ^4$, shows scale anomaly$^{\citen{ber}}$ while the 
addition of a coupling with a Chern-Simons gauge field has been suggested as a
field-theoretical formulation of the Aharonov-Bohm effect$^{\citen{hag}}$ and as
an effective theory for the fractional quantum Hall effect$^{\citen{fra}}$. 

In this paper we discuss the nonrelativistic limit
of the relativistic  theory of a real scalar field with
quartic self-interaction in $2+1$ dimensions. 
This limit is not trivial.
In the corresponding nonrelativistic model$^{\citen
{ber}}$, the self-energy vanishes identically while in the relativistic theory
the lowest order (2 loops) correction is logarithmically divergent. 
On the other hand, the four point function is logarithmically divergent in 
the nonrelativistic model while it is finite in the relativistic case.

To
better illustrate our procedure, we begin by discussing the two
particle scattering.  In Sec. 2, we present the model and calculate the 1PI
four-point function to one loop order in an approximation, for low external
momenta, such that it is possible to know the part of the Hilbert space
each contribution comes from. We obtain the leading correction to the dominant
nonrelativistic particle--particle scattering amplitude, which coincides
with the small $\mid \vec p\mid $ expansion of the exact 1--loop amplitude.
The two-point function is calculated to two loop order in Sec.3 where we
discuss the renormalization of the theory. We present, in Sec.4, a
nonrelativistic reduction scheme for the 2-particle scattering amplitude,
compare our results with those obtained from the corresponding
nonrelativistic model and derive an effective nonrelativistic Lagrangian
that accounts for the results up to order $|\vec p|^2/m^2$.

\section{PARTICLE-PARTICLE AMPLITUDE}

We consider a real self-interacting scalar field in $2+1$ dimensions whose Lagrangian density is
given by 
\begin{equation}
{\cal L}=\frac 12\partial _\mu \phi \partial ^\mu \phi -\frac 12m^2\phi ^2-
\frac \lambda {4!}\phi ^4+{\cal L}{_{c.t.}},  \label{lagran}
\end{equation}
where ${\cal L}{_{c.t.}}$ is the counterterm Lagrangian needed to fix the
mass $m$, the field intensity and the coupling $\lambda $ at their
renormalized values. Our units are such that $\hbar =c=1$ and the Minkowski
metric has signature (1, -1, -1). The mass dimension of
the scalar relativistic field is $1/2$ and that of $\lambda $ is 1.
The quartic self interaction is super-renormalizable and the degree of
superficial divergence of a graph G is given by $d(G)=3-\frac N2-V$, 
where $N$ and $V$ are the numbers of external legs and vertices
respectively.  By assuming a Wick ordering prescription in (\ref{lagran}),
the only divergences are those arising from the two loop
self-energy diagram.

We are going to calculate the two body amplitude to one loop order in an
approximation, for low external momenta, which separates the contributions
coming from the low (L) and high (H) energy intermediary states through the
introduction of an intermediate cutoff $\Lambda$ in the $\vert \vec k \vert$
-integration of the loop momenta. No cutoff is introduced in the integration
over $k^o$. Precisely, the procedure consists of the following steps.
Firstly, we use the Feynman identity 
\begin{equation}
\frac 1{ab^n}=\int_0^1dx\frac{nx^{n-1}}{\left[ (b-a)x+a\right] ^{n+1}}
\label{F}
\end{equation}
and make the necessary change of variables, to put the integrand in a
symmetric form. We then integrate over $k^0$, over the angular part of  $
\vec k$ and perform the parametric integral. The remaining integration over $
\mid \vec k\mid $ is then divided into two parts, corresponding to the low
and high energy contributions of the loop integration, by introducing an
intermediate cutoff $\Lambda $ such that $\mid \vec p\mid <<\Lambda <<m$ and 
$\frac{\mid \vec p\mid }\Lambda \simeq \frac \Lambda m$. In the low energy
sector, $0\leq \mid \vec k\mid <\Lambda $, we approximate the integrand by
expanding it in powers of $\frac{\mid \vec p\mid }m$ and $\frac{\mid \vec k
\mid }m$. In the high energy part, $\Lambda <\mid \vec k\mid <\Lambda
_0(\rightarrow \infty )$, relativistic virtual modes are involved and only $
\frac{\mid \vec p\mid }m$ can be considered a small quantity. Keeping all
contributions up to order $\eta ^2$ , where $\eta \simeq \frac{\mid \vec p
\mid }m$ ( $\simeq \frac{\mid \vec p\mid ^2}{\Lambda ^2}\simeq \frac{\Lambda
^2}{m^2}$ ), we are able to evaluate the amplitude up to order $\frac{\mid 
\vec p\mid ^2}{m^2}$, separating the contributions that come from low and
high loop momenta. It should be noticed that, since in our prescription the
integration over $k^0$ is unrestricted, locality in time is guaranteed.

It should be remarked that the use of Feynman's parameterization (\ref{F})
is not essential. After the $k^0$--integration, one could introduce the intermediate cutoff $\Lambda$, proceed the approximations for the L and H parts of the
loop integration as outlined above, and perform the angular and radial integrations. The possible differences appearing in the outcome of these alternative
calculations are physically irrelevant as we shall see later.

The 1PI four-point function to one loop order is given diagrammatically in
Fig. 1. Notice that the last two diagrams do not appear in the 
nonrelativistic theory where propagation is only forward in time.

Let us initially concentrate in the s-channel amplitude which is given by 
\begin{equation}
A_s(p_1,p_2,m,\lambda )=-\frac{\lambda ^2}2\int \frac{d^3k}{(2\pi )^3}\frac 1
{(k^2-m^2+i\epsilon )((p_1+p_2-k)^2-m^2+i\epsilon )}.  \label{canalS}
\end{equation}
Taking the external momenta on shell and working in the center of mass (CM)
frame (that is, $\vec p_1=-\vec p_2=\vec p$ , $\vec p_1^{^{\phantom{a}\prime
}}=-\vec p _2^{^{\phantom{a}\prime }}=\vec p^{^{\phantom{a}\prime }}$ and $
p_1^0=p_2^0=p_1^{^{\phantom{a}\prime }0}=p_2^{^{\prime }0}=w_p=\sqrt{m^2+
\vec p^2\text{ }}$), this amplitude
can be written, after performing the $k^0$ integration using the
Cauchy-Goursat theorem and the trivial angular integration, as 
\begin{equation}
A_s(\mid \vec p\mid ,\lambda ,m)=\frac{\lambda ^2}{32\pi }\int_0^{\Lambda
_0^2}d(\vec k^2)\frac 1{w_k}\frac 1{\vec p^2-\vec k^2+i\epsilon }  \label{As}
\end{equation}
where $w_k=\sqrt{\vec k^2+m^2}$ and $\Lambda _0$ is an ultraviolet cutoff
that can be made infinity at any time since the graph is finite.  
The above integral can
be calculated exactly and the result expanded for small $\vec p^2$ but, as said before, we
introduce an intermediate cutoff $\Lambda $ to distinguish the contributions
of low and high momenta. 

The low $\vec k^2$ contribution to $A_s$, the integration from 0 to $\Lambda
^2$, is calculated using the approximation 
\begin{equation}
w_k^{-1}= \frac 1m\left[ 1-\frac{\vec k^2}{2m^2}+\frac 38\frac{\vec k^4
}{m^4}-\ldots \right] ,  \label{WkLow}
\end{equation}
and one gets, retaining terms up to ${\cal O}(\eta ^2)$, 
\begin{equation}
A_s^{(L)} \simeq\frac{\lambda ^2}{32\pi m}\left\{ -\left( 1-\frac{\vec p^2}{2m^2
}\right) \left[ \ln \left( \frac{\Lambda ^2}{\vec p^2}\right) +i\pi \right] +
\frac{\vec p^2}{\Lambda ^2}  
 +\frac{\vec p^4}{2\Lambda ^4}+\frac{\Lambda ^2}{2m^2}-\frac{
3\Lambda ^4}{16m^4}\right\} .  \label{AsL}
\end{equation}
Hereafter, the symbol $\simeq$ indicates that the expression which follows 
holds up to order $\eta^2$.

To obtain the high $\vec k^2$ contribution, integrating from $\Lambda ^2$ to 
$\Lambda _0^2$, equation (\ref{WkLow}) can no longer be used but we can
still simplify the integrand by taking 
\begin{equation}
\frac 1{\vec p^2-\vec k^2}= -\frac 1{\vec k^2}\left[ 1+\frac{\vec p^2}{
\vec k^2}+\frac{\vec p^4}{\vec k^4}+\ldots \right] ,  \label{pK}
\end{equation}
resulting, to the same $\eta ^2$ order, 
\begin{equation}
A_s^{(H)} \simeq\frac{\lambda ^2}{32\pi m}\left\{ \left( 1-\frac{\vec p^2}{2m^2}
\right) \ln \left( \frac{\Lambda ^2}{4m^2}\right) -\frac{\vec p^2}{2m^2}-
  \frac{\vec p^2}{\Lambda ^2}-\frac{\vec p^4}{2\Lambda ^4}-\frac{
\Lambda ^2}{2m^2}+\frac{3\Lambda ^4}{16m^4}\right\} .  \label{AsH}
\end{equation}
\noindent
Although we have made distinct approximations (\ref{WkLow}) and (\ref{pK})
in the integrands of the low and high contributions, there exist an exact
cancellation of the $\Lambda $ dependent terms in $A_s^{(L)}$ and $A_s^{(H)}$
, order by order in $\eta $, as it should. Notice that if we had not used the Feynman's trick or
had chosen a distinct routing of the external momenta through the graph, the 
L- and H-parts of the s-channel amplitude would differ from (\ref{AsL}) and (\ref{AsH}) only in the coefficients of the $(|\vec p|/\Lambda)^n$ terms.

The nonrelativistic limit of the amplitude (\ref{As}
), to subleading order, is given by 
\begin{equation}
A_s(\mid \vec p\mid ,\lambda ,m)\simeq-\frac{\lambda ^2}{32\pi m}\left\{
\left( 1-\frac{\vec p^2}{2m^2}\right) \left[ \ln \left( \frac{4m^2}{\vec p^2}
\right) +i\pi \right] +\frac{\vec p^2}{2m^2}\right\} .  \label{AsNR}
\end{equation}
\noindent This is the same result one obtains by evaluating (\ref{As}),
without introducing the cutoff $\Lambda $, and expanding around $\vec p^2=0$.

The t-channel amplitude, in the CM
frame with external momenta on shell, can be written after using Feynman identity (
\ref{F}), making the substitution $k\rightarrow k+qx$ ($q=p-p^{\prime }$
being the transferred momentum), evaluating the $k^0$ and the angular
integrations and integrating in the Feynman parameter, as

\begin{equation}
A_t(\mid \vec p\mid ,\lambda ,m,\theta )=-\frac{\lambda ^2}{32\pi }
\int_0^{\Lambda _0^2}d(\vec k^2)\frac 1{w_k(\vec k^2+\vec q^2/4+m^2)},
\label{Atk}
\end{equation}
where $\vec q^2=2\vec p^2(1-\cos \theta )$ and $\theta $ is the scattering
angle. Proceeding as before, separating the low and high
loop momentum contributions to $A_t$, we get,  
\begin{equation}
A_t^{(L)}\simeq-\frac{\lambda ^2}{32\pi m}\left\{ \frac{\Lambda ^2}{m^2}-\frac{
3\Lambda ^4}{4m^4}\right\}  \label{AtL}
\end{equation}
and 
\begin{equation}
A_t^{(H)}\simeq-\frac{\lambda ^2}{32\pi m}\left\{ 2-\frac{\vec q^2}{6m^2}-\frac{
\Lambda ^2}{m^2}+\frac{3\Lambda ^4}{4m^4}\right\} .  \label{AtH}
\end{equation}
We clearly see that the resultant amplitude, given by 
\begin{equation}
A_t(\mid \vec p\mid ,\lambda ,m,\theta )=-\frac{\lambda ^2}{32\pi m}
\left\{ 2-\frac{\vec p^2}{3m^2}(1-\cos \theta )\right\} ,  \label{AtNR}
\end{equation}
comes entirely from the high-energy states in the Hilbert space while the
s amplitude comes from both low and high momenta. Certainly, this should
be expected for the nonrelativistic limit of a diagram that does not even
exist in the nonrelativistic theory.

The nonrelativistic limit of the u-channel amplitude is
obtained from (\ref{AtNR}) by taking $\theta \rightarrow \theta -\pi $, that
is $A_u(\mid \vec p\mid ,\lambda ,m,\theta )=A_t(\mid \vec p\mid
,\lambda ,m,\theta -\pi )$,  
which corresponds to the exchange of the two particles in the final state.
The sum of these two contributions is independent of the scattering angle and
given by 
\begin{equation}
A_t+A_u\simeq-\frac{\lambda ^2}{32\pi m}\left\{ 4-\frac{2\vec p^2}{3m^2}
\right\} .  \label{AtAu}
\end{equation}
This is the parcel of the one loop amplitude that comes from the diagrams
that involve virtual pair creation and annihilation only allowed in the
relativistic theory. It arises exclusively from the high $\vec k^2$
integration but its contribution is greater than the subleading order of 
(\ref{AsNR}). Although transitions involving relativistic modes have very low
probabilities, the summation of the contributions of all virtual
relativistic momenta turn out to be significant.

Adding the tree amplitude to (\ref{AsNR}) and (\ref{AtAu}), we obtain the
scattering amplitude to one loop order, in our nonrelativistic
approximation, as

\begin{equation}
A_{(1)}(\mid \vec p\mid ,\lambda ,m)\simeq\lambda -\frac{\lambda ^2}{32\pi m}
\left\{ \left( 1-\frac{\vec p^2}{2m^2}\right) \left[ \ln \left( \frac{4m^2}{
\vec p^2}\right) +i\pi \right] +4-\frac{\vec p^2}{6m^2}\right\}
\label{ANRum}
\end{equation}
We see that the leading correction to the dominant term of the low momenta,
nonrelativistic, scattering comes from the high $\vec k^2$ loop integrations
of t- and u- channels, that are intrinsically relativistic. Notice, again,
that the amplitude is finite, though $m>>\mid \vec p\mid $.

The particle-particle amplitude to one loop order can be exactly calculated
in an  arbitrary frame and is given by 
\begin{eqnarray}
A_{(1)} &=&\lambda -\frac{\lambda ^2}{32\pi m} \left\{ \frac 1{\sqrt{(p_1+p_2)^2/4m^2}}\left[ \ln \left( \frac{
\sqrt{(p_1+p_2)^2/4m^2}+1}{\sqrt{(p_1+p_2)^2/4m^2}-1}\right) +i\pi \right]
\right.   \nonumber \\
&&\left. +\left[ \frac 1{\sqrt{(p_1-p_1^{\prime })^2/4m^2}}\ln \left( \frac{
1+\sqrt{(p_1-p_1^{\prime })^2/4m^2}}{1-\sqrt{(p_1-p_1^{\prime })^2/4m^2}}
\right) +(p_1^{\prime }\rightarrow p_2^{\prime })\;\right] \right\} , 
\label{AumEG}
\end{eqnarray}
where $p_i$ and $p_i^{\prime }$ are the external on shell incoming and outgoing
momenta. Taking this expression in the CM frame and expanding for small 
$|\vec p_i|^2$, it reduces to (\ref{ANRum}), showing that our
approximation reproduces the small momentum expansion of the exact result up
to order $\vec p^2/m^2$. Therefore, although apparently unnecessary in the $
\phi ^4$ theory where one has exact results, our approximation procedure can
be applied with confidence in other theories where, eventually, exact
analytical calculations can not be done. It should also be noticed that the
separation of the contributions of low and high energy intermediate states
to the amplitude can be effectuated in the general situation but, for
simplicity, we made it in the CM frame.

\section{PARTICLE SELF-ENERGY}

With normal ordering imposed to the Lagrangian (\ref{lagran}), the first
nonvanishing contribution for the two point function comes from the two loop
diagram of Fig.2.
This is the unique primitive divergent diagram of $\lambda \phi ^4$ theory
in 2 + 1 dimensions. This type of diagram and, actually, all the
self-energy insertions are not allowed in the nonrelativistic theory where
the propagation is always forward in time. There, the physical mass is the
natural parameter and the full propagator coincides with the free one. We
shall calculate this two loop self-energy exactly and also by the same
procedure of separating low and a high internal momentum contributions, as
we did before. In this case, however, the ultraviolet cutoff $\Lambda _0$
can not be made infinity before a subtraction because the graph is
logarithmically divergent. We shall see that both the finite part and the
divergent one come from the region of integration where both loop momenta
are high.

Before trying to calculate any of the $k$ or $l$ integrations we completely
disentangle these variables by proceeding as follows. We apply (\ref{F}) to
the $k$-loop and make the substitution $k\rightarrow k+lx$. We then repeat 
(\ref{F}), rescale the variables such that $k\rightarrow k/\sqrt{y}$ and $
l\rightarrow l/\sqrt{C}$ and perform the translation $l\rightarrow l+\frac{
(1-y)}{\sqrt{C}}p$ to obtain 
\begin{equation}
\stackrel{\wedge }{\Sigma }_{(2)}(p,m,\lambda ,\Lambda _0) =\frac{i\lambda
^2}{192\pi ^6}\int_0^1dx\int_0^1dy\frac y{\left[ yC\right] ^{3/2}}  
\int d^3kd^3l\frac 1{\left[ k^2+l^2+Dp^2-m^2+i\epsilon \right] ^3}
\label{SE}
\end{equation}
\noindent
where $C=C(x,y)=yx(1-x)+(1-y)$ and $D=D(x,y)=(1-y)\left(
1-(1-y)/C(x,y)\right)$.  It can be shown that these functions satisfy $0\leq C\leq 1$ and $ 0\leq D\leq 1/9$ for all $0\leq
x,y\leq 1$.

Owning to the form of the integrand of (\ref{SE}), the $(k^0,l^0)$
integration can be exactly done using polar coordinate with $r=\sqrt{{k^0}^2+
{l^0}^2}$ and $\alpha =\arctan \left( \frac{l^0}{k^0}\right) $ resulting,
after trivial angular integration, in 
\begin{equation}
\stackrel{\wedge }{\Sigma }_{(2)}(p,m,\lambda ,\Lambda _0)=\frac{i\lambda ^2
}{384\pi ^3}\int_0^1dx\int_0^1dy\frac y{\left[ yC\right] ^{3/2}}
F(p^2,m^2,x,y,\Lambda _0)  \label{SEkl}
\end{equation}
\noindent
where 
\begin{equation}
F(p^2,m^2,x,y,\Lambda _0)=\int_0^{\Lambda _0^2}d(\vec k^2)\int_0^{\Lambda
_0^2}d(\vec l^2)\frac 1{\left[ \vec k^2+\vec l^2+m^2-p^2D(x,y)-i\epsilon
\right] ^2}\,.  \label{Fxykl}
\end{equation}

The $\vec k^2$ and $\vec l^2$ integrations above can be exactly performed
giving, for $\Lambda _0>>m,$

\begin{equation}
F(p^2,m^2,x,y,\Lambda _0)=\ln \left( \frac{\Lambda _0^2}{m^2}\right) -\ln
\left[ 2\left( 1-\frac{p^2}{m^2}D(x,y)\right) \right] .  \label{Fxy}
\end{equation}
\noindent
To see from where this contribution comes, as we did before, we introduce
the same intermediate cutoff $\Lambda ^2$ for both $\vec k^2$ and $\vec l^2$
integrations dividing the $(\vec k^2,\vec l^2)$ quadrant into four parts
denoted, in a self-explained notation, as $L-L$, $L-H$, $H-L$ and $H-H$. We
then obtain the contributions for $F$ coming from each of these parts as 
\begin{eqnarray}
F_{L-L} &\simeq &\frac{\Lambda ^4}{b^2m^4}  \label{FLL} \\
F_{L-H} &=&F_{H-L}\simeq \frac{\Lambda ^2}{bm^2}-\frac{3\Lambda ^4}{2b^2m^4}
\label{FLH} \\
F_{H-H} &\simeq &-\frac{2\Lambda ^2}{bm^2}+\frac{2\Lambda ^4}{b^2m^4}-\ln
2-\ln b+\ln \left( \frac{\Lambda _0^2}{m^2}\right) ,  \label{FHH}
\end{eqnarray}
\noindent
where $b(p^2/m^2,x,y)=1-D(x,y)\,p^2/m^2-i\epsilon$. This clearly shows that $\stackrel{\wedge }{\Sigma }_{(2)}$ comes entirely
from the high, relativistic, virtual states, as it is expected.

Inserting (\ref{Fxy}) in (\ref{SEkl}), the cutoff regulated two loop self-energy, is given by 
\begin{equation}
\stackrel{\wedge }{\Sigma }_{(2)}(p,m,\lambda ,\Lambda _0)=\frac{i\lambda ^2
}{192\pi ^2}\left[ \ln \left( \frac{\Lambda _0^2}{m^2}\right) -E\left( \frac{
p^2}{m^2}\right) \right] ,  \label{SEreg}
\end{equation}
\noindent
where the function $E(z)$ is defined by 
\begin{equation}
E(z)=\frac 1{2\pi }\int_0^1dx\int_0^1dy\frac{y\ln \left( 2\left[
1-zD(x,y)\right] \right) }{\left[ yC(x,y)\right] ^{3/2}}.  \label{E}
\end{equation}
\noindent
Notice that, the procedure described just above (\ref{SE}) can be readily
extended to any dimension yielding a much easier computation of the
``sunset''graph of Fig.2, in the general case. Also, if one treats (\ref{SE}) 
by dimensional regularization, one gets

\begin{equation}
\stackrel{\wedge }{\Sigma }_{(2)}^{(\dim )}(p,m,\lambda ,d)=\frac{i\lambda ^2
}{192\pi ^2}\left[ \frac 1{3-d}+\left( \ln 2-\gamma \right) -E\left( \frac{
p^2}{m^2}\right) \right] ,  \label{SEdim}
\end{equation}
\noindent so that by making a minimal subtraction, the finite part obtained
differs from that of (\ref{SEreg}) by a constant term, as it should.

The mass and wave function renormalization program can now be implemented
. The full propagator is given by $G_R(p^2)=i(p^2-m^2-i\Sigma )^{-1}$, where 
$\Sigma =\stackrel{\wedge }{\Sigma }+i(Z-1)(p^2-m^2)-i\delta m^2Z$. Particle interpretation of the theory requires that the complete propagator $
G_R$ has a pole of residue $i$ at $p^2=m^2$ which implies that 
\begin{equation}  \label{dm}
\delta m^2=m^2\left( \frac{\lambda ^2}{192\pi ^2m^2}\right) \left[\ln \left( \frac{\Lambda _0^2}{m^2}\right)-E(1)\right]   \label{deltam2}
\end{equation}
\noindent
and 
\begin{equation}
Z=1+\left( \frac{\lambda ^2}{192\pi ^2m^2}\right) \left. \frac{\partial E}{
\partial z}\right| _{z=1},  \label{Z}
\end{equation}

\noindent
up to $\lambda ^2$ order. As we saw in the last section, the four-point
function is finite so that no coupling constant renormalization is necessary.

\section{NONRELATIVISTIC REDUCTION}

The approximation we have used, introducing an intermediate cutoff $\Lambda $
, not only permits the identification of the origin (in the Hilbert space)
of each contribution, but it also allows the construction of a
nonrelativistic reduction scheme at the level of the Green's functions.

Adding separately the low and high energy contributions of each channel to
the scattering amplitude, in an arbitrary reference frame but for
external nonrelativistic particles on the mass shell, one obtains (up to
order $\frac{\mid \vec p\mid ^2}{m^2}$ )

\begin{eqnarray}
A_{(1)}^{(L)} &\simeq&\lambda +\frac{\lambda ^2}{32\pi m}\left\{ -\left( 1-\frac{(
\vec p_1-\vec p_2)^2}{8m^2}\right) \left[ \ln \left( \frac{4\Lambda ^2}{(
\vec p_1-\vec p_2)^2}\right) +i\pi \right] \right.  \nonumber \\
&&\ \left. +\frac{(\vec p_1-\vec p_2)^2}{4\Lambda ^2}+\frac{(\vec p_1-\vec p
_2)^4}{8\Lambda ^4}-\frac{3\Lambda ^2}{2m^2}+\frac{21\Lambda ^4}{16m^4}
\right\}  \label{ANRL}
\end{eqnarray}
and

\begin{eqnarray}
A_{(1)}^{(H)} &\simeq&\frac{\lambda ^2}{32\pi m}\left\{ \left( 1-\frac{(\vec p_1-
\vec p_2)^2}{8m^2}\right) \ln \left( \frac{\Lambda ^2}{4m^2}\right) -\frac{(
\vec p_1-\vec p_2)^2}{4\Lambda ^2}\right.  \nonumber \\
&&\ \ \left. -\frac{(\vec p_1-\vec p_2)^4}{8\Lambda ^4}+\frac{3\Lambda ^2}{
2m^2}-\frac{21\Lambda ^4}{16m^4}-4+\frac{(\vec p_1-\vec p_2)^2}{24m^2}
\right\} .  \label{ANRH}
\end{eqnarray}
\noindent 
One should naturally expects that the low $|\vec k|$ contribution expresses, to some extent, the scattering amplitude
obtained from a nonrelativistic (NR) theory although the arbitrariness in the
introduction of the intermediate cutoff prevents any straightforward
identification.

We must note that these amplitudes were calculated from a relativistic
theory in which the states are normalized as $\langle \vec p^{\phantom{a} 
\prime }|\vec p\rangle =2w_p\delta ^3(\vec p^{\phantom{a}\prime}-\vec p)$. 
On the other hand, the usual
normalization of states in a nonrelativistic theory does not have the
$2w$ factor. Thus, for the purpose of comparison, it is necessary to multiply
our results by
\begin{equation}
(16w_{p1}w_{p2}w_{p1^{\prime }}w_{p2^{\prime }})^{-1/2}= \frac 1{4m^2}
\left( 1-\frac{\vec p_1^2+\vec p_2^2}{2m^2}+...\right)   \label{NF}
\end{equation}

Let us initially analyze the above expressions for the amplitude up to
the dominant order of the 1-loop correction, that is, let us consider

\begin{equation}
A^{(L)}= \frac{A_{(1)}^{(L)}}{4m^2}=\frac{\lambda}{4m^2} -\frac{\lambda ^2}{
128\pi m^3}\left[ \ln \left( \frac{4\Lambda ^2}{(\vec p_1-\vec p_2)^2}
\right) +i\pi \right]  \label{ALdom}
\end{equation}
and

\begin{equation}
A^{(H)}=\frac{A_{(1)}^{(H)}}{4m^2}=\frac{\lambda ^2}{128\pi m^3}\ln \left( \frac{
\Lambda ^2}{4m^2}\right) .  \label{AHdom}
\end{equation}
One can see that equation (\ref{ALdom}) coincides with the result from the
nonrelativistic theory specified by the Lagrangian density

\begin{equation}
{\cal L}^{NR}=\phi ^{\dagger }\left( i\partial _t+\frac{\nabla ^2}{2m}
\right) \phi -\frac{v_0}4(\phi ^{\dagger }\phi )^2,  \label{lagranNR}
\end{equation}
\noindent with $v_o=\lambda /4m^2$\ (compare with equation (2.13) of ref.
\citen{ber}), if $\Lambda $ is reinterpreted as a genuine 
ultraviolet cutoff. Such an interpretation, however, can only be sustained
after performing a nonrelativistic reduction procedure as follows. First,
notice that, neglecting terms of order $\eta\simeq |\vec p|/m $ or higher in the self-energy,
one gets $\stackrel{\wedge }{\Sigma }_{(2)L-L}=0$ ($=\stackrel{\wedge }{
\Sigma }_{(2)L-H}$) showing that in this case the low energy contribution
for $\stackrel{\wedge }{\Sigma }_{(2)}$ vanishes identically. This
agrees with the nonrelativistic result where there is no radiative correction at all
to the propagator. We then fix the parameter $m$ and promote $\Lambda $
to be the ultraviolet cutoff of the reduced nonrelativistic theory. This
last step is the fundamental reinterpretation required for our reduction
process. It produces an unrenormalized logarithmic divergent four-point
function as one has in the nonrelativistic theory (\ref{lagranNR}).

The above nonrelativistic reduction of the leading term of the
L-contribution to the two particle scattering amplitude is equivalent to the 
$m\rightarrow \infty $ limit effectuated on the classical Lagrangian$^{\citen{ber},\citen{jac}}$ and it is also reproduced by making a Foldy-Wouthuysen
transformation in the free part of the Lagrangian (\ref{lagran})$^{\citen{bd}}$.
An interesting aspect of this reduction procedure is that the contribution
of high energy states appears providing the necessary counterterm to make
the amplitude finite instead of logarithmic divergent. The divergence
produced in the low energy contribution (\ref{ALdom}) would be naturally
compensated by the high part, if the full, relativistic, theory were
considered. One can, in this way, better understand the renormalization of
the nonrelativistic model of ref. \citen{ber}.

It is worthwhile to mention that, one could naively think that the
divergence of the 2-particle amplitude calculated with the nonrelativistic
theory is due to the complete exclusion of the propagation backwards in
time. In fact, if one splits the relativistic propagator as a sum of its
positive (particle) and negative (antiparticle) frequency parts,

\begin{equation}
\frac i{k^2-m^2+i\epsilon }=\frac 1{2w_k}\frac i{k^0-w_k+i\epsilon }+\frac 1{
-2w_k}\frac i{k^0+w_k+i\epsilon }\quad ,  \label{prel}
\end{equation}
the relativistic CM s-channel amplitude can be written as $
A_s=A_s^{+}+A_s^{-}$ where 

\[
A_s^{\pm }=\frac{\lambda ^2}{32\pi }\int_0^{\Lambda _0^2}d(\vec k^2)\frac 1{
 2w_{k}^{2}}\frac 1{w_p\mp w_k+i\epsilon }\quad ,
\]
with the superscripts $+$ and $-$ denoting the contributions of the
particle-particle and the antiparticle-antiparticle propagations,
respectively. As one immediately sees, both of these parcels are finite when 
$\Lambda _0\rightarrow \infty .$ Naturally, one can introduce an auxiliary
cutoff to separate the L- and the H- contributions for each part, and doing
so, one finds that the particle propagation contribution is dominant, as
expected.

Let us now examine the sub-dominant order. Disregarding constant terms which
can be absorbed into a coupling constant renormalization, the low energy
part is 
\begin{eqnarray}
A^{(L)} &=&\frac \lambda {4m^2}-\frac \lambda {4m^2}\frac{(\vec p_1+\vec p_2)^2+(
\vec p_1-\vec p_2)^2}{4m^2}  \nonumber \\
\phantom a &&-\frac{\lambda ^2}{128\pi m^3}\left[ 1-\frac{2(\vec p_1+\vec p
_2)^2+3(\vec p_1-\vec p_2)^2}{8m^2}\right] \left[ \ln \left( \frac{4\Lambda
^2}{(\vec p_1-\vec p_2)^2}\right) +i\pi \right] .\label{LL}
\end{eqnarray}
\noindent
To reproduce the new terms appearing in this expression, we add to (\ref
{lagranNR}) the effective interaction Lagrangian 
\begin{equation}
{\cal L}_{int}^{NR}=\frac{v_1}4\left( \phi ^{\dagger }\frac{(\nabla ^2\phi
^{\dagger })}{m^2}\phi ^2+\frac{(\vec \nabla \phi ^{\dagger })^2}{m^2}\phi
^2\right) +\frac{v_2}4\left( \phi ^{\dagger }\frac{(\nabla ^2\phi
^{\dagger })}{m^2}\phi ^2-\frac{(\vec \nabla \phi ^{\dagger })^2}{m^2}\phi
^2\right) 
\end{equation}
which is the more general, dimension 6, quadrilinear local
nonrelativistic interaction. For the calculation of the contributions
arising from these new vertices we will have to introduce ultraviolet
cutoffs. It is easily verified that the polynomial part of the result
is cutoff dependent and this freedom can be used to adjust it to match
the polynomial part of (\ref{LL}). For that reason, we restrict the
discussion to the non-polynomial part of the additional contribution
which, in one loop order and up to ${\cal O}(\vec p^2/m^2)$, is
(again, disregarding constant terms)
  
\begin{equation}
-\frac m{8\pi }v_0\left( v_1\frac{(\vec p_1+\vec p_2)^2}{m^2}+v_2\frac{(\vec p
_1-\vec p_2)^2}{m^2}\right) \left[ \ln \left( \frac{4\Lambda ^2}{(\vec p_1-
\vec p_2)^2}\right) +i\pi \right]. 
\end{equation}
Comparing with (\ref{LL}) we find
\begin{equation}
v_1= -\frac{\lambda}{16m^2} \qquad \mbox{and} \qquad v_2=
-\frac{3\lambda}{32m^2},
\end{equation}
which fixes the effective nonrelativistic Lagrangian up to the order
$\vec p^2/m^2$. As before, the high energy part furnishes theo counterterm needed to make the amplitude finite.

\section{Conclusions}

In this work we discussed the nonrelativistic limit of the 2+1
dimensional $\phi^4$ theory by means of a scheme which separates the
contributions from the high and low virtual momenta in the loop
integrations. This method provides a systematic way of extracting
different orders in $\vert \vec p\vert/m$ in the nonrelativistic
approximation and can be applied to more general situations.
Proceeding along these lines, we were able to derive an
effective Lagrangian which, up to order $\vert \vec p^2\vert/m^2$,
correctly reproduces the nonrelativistic limit. The interaction
Lagrangian so obtained is equivalent, in the leading order, to the
quantum mechanical delta function potential. The new terms arising in
the subleading order, however, can not be interpreted in terms of a
two body potential.

\end{document}